\documentclass[%
 reprint,
 amsmath,amssymb,
 aps,nofootinbib,
]{revtex4-1}
\usepackage{bm}
\PassOptionsToPackage{linktocpage}{hyperref}
\usepackage[hyperindex,breaklinks]{hyperref}
\usepackage{enumitem}
\usepackage{slashed}

\renewcommand{\theta}{\vartheta}

\newcommand{\ket}[1]{\ensuremath{\left|\, #1\right>}}

\usepackage{array}
\usepackage{mathtools}

\usepackage{etoolbox}

\begin{document} 

\title{Area Law  Micro-State Entropy from Criticality and 
Spherical Symmetry}

\author{Gia Dvali} 
\affiliation{%
Arnold Sommerfeld Center, Ludwig-Maximilians-Universit\"at, Theresienstra{\ss}e 37, 80333 M\"unchen, Germany, 
}%
 \affiliation{%
Max-Planck-Institut f\"ur Physik, F\"ohringer Ring 6, 80805 M\"unchen, Germany
}%
 \affiliation{%
Center for Cosmology and Particle Physics, Department of Physics, New York University, 726 Broadway, New York, NY 10003, USA
}%

\date{\today}

\begin{abstract}
It is often assumed that the area law of micro-state entropy and 
the holography are intrinsic properties exclusively of the gravitational systems, such as black holes. We construct  a non-gravitational model that exhibits an entropy that scales as area of  
a sphere of one dimension less. It is represented by a  non-relativistic bosonic field living on a $d$-dimensional sphere of radius $R$ and experiencing an angular-momentum-dependent attractive interaction.  We show that the system possesses a quantum critical point with the emergent gapless modes. Their number is equal to the area of a $d-1$-dimensional sphere of the same radius $R$. 
These gapless modes create an exponentially large number of degenerate micro-states 
with the corresponding micro-state entropy given by the area of the same $d-1$-dimensional sphere.  Thanks to a double-scaling limit, the 
counting of the entropy and of the number of the gapless modes is made exact. 
The phenomenon takes place for arbitrary number of dimensions 
and can be viewed as a version of {holography}. 
\end{abstract}


\maketitle

\section{Introduction}
 Since the celebrated paper by Bekenstein \cite{Bek}, one of the main open problems in gravity  is to understand the origin of the area law for the black hole entropy. 
 This law is often viewed as a defining property of black holes, something that is linked with the most fundamental aspect of quantum gravity.\\
 
 The area scaling of the black hole entropy gave rise to several hypothesis,  the most well known of which goes under the name of {\it holography} \cite{Hologram}.  
 Another famous conjecture revealing the holographic nature of gravity, is 
 AdS/CFT correspondence \cite{ADS1,ADS2,ADS3}. Once again, it appears that  information of a gravitational system can be summed up in 
the degrees freedom of a lower dimensional theory \cite{ADS3, ADSH}. \\ 

In the light of the above developments it is important to understand whether the area law entropy and holography are fundamentally linked to gravity, or represent a more general phenomenon of nature, which could take place also in non-gravitational systems. This point of view was advocated  in \cite{DG}. \\

For shedding some light on the above question, it would be 
extremely useful to have a simple quantum model - even if at a toy 
level - in which the area law scaling of entropy could be unambiguously   
demonstrated, without being blurred by the uncontrollable quantum gravity corrections.  We are not aware of any such model in the literature and it is the purpose of the present paper to construct one. \\

 Before continuing, we would like to mention an important work \cite{SV}, which gives a microscopic counting of entropy 
 for the extremal black holes in string theory, using the power of supersymmetry.  The attitude that we shall take is much less profound and does not rely on any of these properties. Of course, we shall only present a toy model, but the fact that it gives both the area-type law and holography - 
without any obvious involvement of gravity - is somewhat remarkable.  \\

 Therefore, in this note we shall attempt to construct an example of a simple non-gravitational 
 quantum system which exhibits  the emergence of the gapless modes with their number scaling as the area of a lower dimensional sphere.  \\
 
  Some clarifications of terminology are in order: \\
 
  1) First, under the term {\it entropy} we do not mean the entanglement entropy, which usually scales as the area. 
 Rather, we mean the true {\it micro-state entropy}, i.e.,  the log from the number of the degenerate micro-states. \\
 
 2) Secondly, we use the term {\it area law} despite the fact that 
the manifold that our field lives on is a $d$-sphere and has no space boundary.  However, this term indicates that the entropy scales not as the volume of a $d$-sphere, but as the volume of $(d-1)$-sphere of the same radius, or equivalently, as the area of a boundary of the 
$d$-dimensional hemisphere. 
 \\

 3) Thirdly, the term {\it non-gravitational} is used  solely due to the external appearance of the model, which does not seem to contain any massless spin-2 field. It may very well be
 (and we think this is indeed the case)  that the model secretly captures some key properties of the black hole gravity, along the ideas of
 \cite{DG}.    
 \\

 The model that  we shall construct,  describes a quantum mechanical bosonic field, $\hat{\psi}$, living in a space of a $d$-dimensional sphere of radius $R$. 
We show that this  system exhibits a quantum critical point where many gapless modes emerge.   Their number is given by the area of a $d-1$-dimensional sphere 
of the same radius $R$. Correspondingly, the entropy of the resulting micro-states 
scales as the area of the $d-1$-dimensional sphere as well. \\

  As we shall see, the area scaling of the micro-state entropy is the result of the interplay of the two
 effects. These are: The quantum criticality of attractive bosons  that creates large number of gapless modes, and the level-degeneracy  due to spherical symmetry that ensures that the number of gapless modes scales as area.  
   Without both of these effects in place, the system would result neither in the area law
   nor in holography.  \\

 The presented model provides a supportive evidence for the hypothesis 
 of \cite{DG}, according to which the origin of the black hole entropy 
 and holography lies in {\it quantum criticality}  of attractive bosons (gravitons).   
 It shares some similarities with the model discussed earlier 
 in \cite{goldstone}, but exhibits some key  differences. 
 We shall first give a detailed description of the model and later 
 give the interpretation of our results. \\
 
\section{Model} 

 The Hamiltonian of the model has the following form: 
   \begin{equation} 
    \hat{H}  =  - \int d^d\Omega \,\, \hat{\psi}^{\dagger} \Delta  \hat{\psi}\,  +  \, g\Omega \,  (\hat{\psi}^{\dagger}  \Delta  \hat{\psi}^{\dagger}) (\hat{\psi} \Delta  \hat{\psi}) \, .
 \label{H1} 
 \end{equation}  
  Here, $d^d\Omega$ is a volume element of a $d$-dimensional unit sphere, 
  with the angular coordinates $\theta_a \,\, (a=1,...d)$ and 
  the total volume $\Omega$. 
   The operator  $\Delta \equiv {\hbar^2  \over 2mR^2}\Delta_d$ is a rescaled Laplace operator, with $\Delta_d$ being the usual covariant  
  Laplace operator on a $d$-dimensional unit sphere.  
 $R$ is the radius of the proper sphere.  
  The parameter $m$ is the mass of the boson 
  and $g$ is a positive coupling constant. Note, $\Delta$ 
  has the dimensionality of energy and correspondingly  $g$ has  the dimensionality of  inverse energy. \\
    
   The operator $\hat{\psi}(\theta_a)$ is a bosonic field that describes a  distribution of the particle number on the  $d$-sphere.   We can represent it as an infinite sum over   the creation and annihilation operators for the modes with different generalized angular momenta, 
  \begin{equation} 
      \hat{\psi} = \sum_{k} \, Y_k(\theta_a) \hat{a}_k \,, 
     \label{expansion} 
   \end{equation}   
 where $Y_k(\theta_a)$ are the spherical harmonic functions on $S_d$. The
 label $k$ stands for a set of $d$ integers
 $k \equiv (k_1,...,k_d)$, which satisfy
 $|k_1|\leqslant k_2 \leqslant ...\leqslant k_d = 0,1,...,\infty$. 
 That is, $Y_k \equiv Y_{k_1,...,k_d}$. 
 These spherical harmonics form a complete orthonormal set. Thus,  among other properties,  they satisfy: 
 \begin{equation} 
 \int d\Omega  \, Y_k^*Y_{k'} = \delta_{kk'} \, ,
  \label{ort} 
  \end{equation} 
  where, $\delta_{kk'} \equiv \delta_{k_1k_1'}\delta_{k_2k_2'}...\delta_{k_dk_d'}$. 
  They represent the eigenfunctions of the rescaled  
 Laplace operator 
   \begin{eqnarray}  \label{eigen} 
 &&  \Delta  Y_k\,   = - \epsilon_k  Y_{k} \,, \\  
 &&  {\rm where}, \,~  
 \epsilon_k \equiv {\hbar^2  \over 2mR^2} k_d(k_d + d-1) \,. \nonumber
   \end{eqnarray}
 The eigenvalues $\epsilon_k$ have the dimensionalities of energy, 
 measured in units of the quantity ${\hbar^2  \over 2mR^2}$. 
It is very important for what follows that the level $\epsilon_k$ exhibits the following  degeneracy, 
\begin{equation}
 {\mathcal N}_k =  \sum_{k_{d-1}=0}^{k_{d} }  \sum_{k_{d-2}=0}^{k_{d-1}}...\sum^{k_2}_{k_{1} = - k_2} \sim  (k_d)^{d-1}\,.
  \label{deg}
  \end{equation} 
  
 The operators $\hat{a}_k^{\dagger}, \hat{a}_k$, are the 
 creation and annihilation operators for the eigenmodes
 corresponding to the eigen-number-sets $k$. They satisfy the usual    
 algebra, 
       \begin{equation} 
    [\hat{a}_j,\hat{a}_k^{\dagger}] = \delta_{jk}\,, \, \, 
  [\hat{a}_j,\hat{a}_k]  =   [\hat{a}_j^{\dagger},\hat{a}_k^{\dagger}] =0\,.  
    \label{algebra} 
 \end{equation} 
 
 Before continuing, we must note that the Hamiltonian  (\ref{H1}) looks unbounded from below.  However, this is not a problem.  We must  assume that the higher order 
stabilizing terms are included.
  As we shall show explicitly, such terms do not change the essence of the phenomenon. Therefore, we shall postpone the discussion of a Hamiltonian with the stabilizing terms included till later. 
For the time-being, we shall ensure the boundedness of the energy    by placing a cutoff on the maximal value of the angular momentum. 
Since the hamiltonian  (\ref{H1}) is particle number conserving, 
the angular momentum cutoff ensures boundedness of energy for any finite number of quanta.  Later, we shall allow the 
particle number to be infinite, but in a special limit, which ensures 
that the energy is bounded from below. 
 \\
 
Let us now rewrite the Hamiltonian in the language of the angular momentum modes. 
 Integrating over the angular coordinates,  we rewrite Hamiltonian in terms 
 of the creation and annihilation operators, 
\begin{eqnarray} \label{HA} 
 \hat{H}  &&= \sum_k \epsilon_k \hat{a}_k^{\dagger}\hat{a}_k - \\
  - g \Omega&&\sum_{s,k,q,r} C_{skqr}  \epsilon_k  \epsilon_r \hat{a}_s^{\dagger}\hat{a}_k^{\dagger}\hat{a}_q \hat{a}_r\,, \nonumber 
 \end{eqnarray}
where we have introduced a notation: $ C_{skqr}  \equiv \int d^d\Omega Y_{s}^* Y_{k}^*Y_{q}Y_{r}$.
\\

 We shall now impose a cutoff that limits the maximal value of the angular momentum, $k_d \leqslant k_{*}$.  Then, 
 according to the definition (\ref{eigen}), the corresponding 
eigenvalue of the Laplacian,  is given by:   
 \begin{equation}
 \epsilon_* \equiv {\hbar^2  \over 2mR^2} k_*(k_* + d-1) \,.
\label{cutoff} 
 \end{equation} 
 Notice, our main result is insensitive to the precise value of $k_*$, as long as it is much larger than one.  Equivalently, the only
natural assumption we are making is that the cutoff level $\epsilon_*$  is much higher than the elementary level ${\hbar^2  \over 2mR^2}$. 
Its precise value is unimportant for our analysis. \\

 Next, let us focus our attention onto the states in which only 
 the $k=0$ mode is macroscopically occupied.  That is, the  corresponding occupation number is very high,  $\langle \hat{a}_0^{\dagger} \hat{a}_0 \rangle  = N_0 \gg 1$.  
In the same time, we restrict the occupation numbers of $k \neq 0$  modes to be much smaller.  On such states, we can use the Bogoliubov approximation \cite{bogoliubov} and replace 
 the operators  $\hat{a}_0^{\dagger},  \hat{a}_0$ by the 
 $c$-numbers,  $\hat{a}_0^{\dagger} = {\rm e}^{- i\alpha} \sqrt{N_0} \,,~
 \hat{a}_0 = {\rm e}^{i\alpha} \sqrt{N_0}$, where  $\alpha$ is an unimportant phase.  \\
 
  In order to reveal the existence of a highly degenerate  
quantum critical point, we shall take the following useful limit.  Notice, for a fixed value of $k_{*}$, we are left with two independent 
  control parameters: The occupation number $N_0$ and the coupling strength $g$.  Using these two control parameters, we shall take the following {\it double-scaling}  limit: 
  \begin{equation} 
 N_0 \rightarrow \infty\,, ~ ~  g  \rightarrow 0 \,,  
 ~~ gN_0 = {\rm finite} \,.
 \label{doublescale}
 \end{equation} 
For the time being, we shall keep $(gN_0)$ as the remaining free control parameter. \\

In order to understand what happens with the interactions terms in the limit
(\ref{doublescale}), let us split the sum into the pieces containing  two, one 
 or none of the zero-mode operators $\hat{a}_0^{\dagger},  \hat{a}_0$
 respectively. 
 Notice, none of the terms can contain more than two zero-mode operators, because they carry zero angular momentum. Thus, the second sum 
in (\ref{HA})  can be rewritten in the following form,  
  \begin{eqnarray}  \label{terms} 
  && {\rm Int.~terms} = \, - g \hat{a}_0^{\dagger}\hat{a}_0 \sum_{k\neq 0} \epsilon_k^2  \, \hat{a}_k^{\dagger}\hat{a}_k - \\
  - && g \hat{a}_0 \Omega\sum_{s,k,r \neq 0} C_{sk0r}  \, \epsilon_k \epsilon_r \, \hat{a}_s^{\dagger} \hat{a}_k^{\dagger}\hat{a}_r \, \nonumber  \\ 
  - && g \hat{a}^{\dagger}_0 \Omega\sum_{k,q,r \neq 0} C_{0kqr} \, \epsilon_k \epsilon_r \,  \hat{a}_k^{\dagger}\hat{a}_q \hat{a}_r \, \nonumber  \\ 
 - && g \Omega\sum_{s,k,q,r \neq0} C_{skqr} \, 
 \epsilon_k \epsilon_r \, \hat{a}_s^{\dagger}\hat{a}_k^{\dagger}\hat{a}_q \hat{a}_r\,, \nonumber  
\end{eqnarray}
where in the fist term we have used the property 
$C_{0k0r} = {\delta_{kr} \over \Omega}$.  The sum is taken
 up to $k_{*}$. \\
  
Now, in the double-scaling limit  (\ref{doublescale}) the last three terms vanish. Indeed, the second and the third terms scale as ${1 \over \sqrt{N_0}}$, whereas the last term scales as ${1 \over N_0}$.  Therefore, the  only surviving term is the first one. Thus, in the double-scaling limit
(\ref{doublescale}), the Hamiltonian (\ref{HA}) 
takes the  following simple form: 
 \begin{equation}  \label{HAL} 
 \hat{H}  =   \sum_{k\neq 0}^{k_{*}}  \epsilon_k \left(
 1 - g N_0 \epsilon_k \right) \hat{a}_k^{\dagger}\hat{a}_k .
 \end{equation}
 We shall now use the remaining control parameter $(N_0g)$ and take the 
  limit,
  \begin{equation} 
    g N_0 \rightarrow  {1 \over \epsilon_*} \,, 
    \label{KMAXlimit} 
\end{equation} 
  but in such a way that the quantity 
  \begin{equation}
  \label{smalldelta}
  \delta_* \equiv \epsilon_* \left(
 1 - g N_0 \epsilon_* \right)  
 \end{equation} 
  approaches zero from the positive side.
 Obviously, $\delta_*$ represents the energy gap of the level-$k_*$ modes. 

 The resulting Hamiltonian is,  
  \begin{equation}  \label{HALSTAR} 
 \hat{H}  =   \sum_{k\neq 0}^{k_{*}}  \epsilon_k \left(
 1 - {\epsilon_k \over \epsilon_*} \right ) 
 \hat{a}_k^{\dagger}\hat{a}_k\,  .
 \end{equation}
 Thus, we arrive to a one-parameter family of the theories labeled 
 by  $\epsilon_*$. 
  \\

  Let us examine the resulting spectrum of the theory. 
  First of all, all the modes with
  $k = k_{*}$ become exactly {\it gapless}. The number of such modes 
  is equal to the degeneracy of the eigenvalue  $k_d = k_{*}$. 
   According to (\ref{deg}) this degeneracy is given  by ${\mathcal N}_{k^*}  \sim k_*^{d-1}$ and thus scales as the area of a $d-1$-dimensional sphere.
   So does the number of the gapless modes:  
   \begin{equation} 
      N_{modes} \, \sim \, \left ( {R \over L_{*}} \right )^{d-1} \, , 
   \label{Nmodes}
   \end{equation} 
   where $L_* \equiv {\hbar \over \sqrt{2m\epsilon_*}}$. 
   That is, if we fix the energy level $\epsilon_*$ of the critical angular 
 excitation and vary the radius of the $d$-sphere, the number of gapless 
   modes that can fit in it, scales as $R^{d-1}$.  
  Note, the choice of the scale $L_*$ is somewhat analogous to the choice 
  of the Planck length in gravity.    \\

   In the same time, the modes with $k_d < k_{*}$, have positive energy gaps. For example,  the modes with $k_d =  k_{*} -1$, have the energy gaps of order $\sim {\hbar^2 \over mR^2}  k_{*}$, which is much higher than the 
 unit energy gap. Thus, the modes $k < k_{*}$ 
 do not contribute into the micro-state entropy of the system. \\ 
  
   Hence, the dominant contribution to the micro-state entropy is coming from the gapless
   modes, with $k = k_{*}$. 
   By ``mildly" populating these modes, we can create an exponentially large 
   number of micro-states. For example, the number of basic 
   micro-states in which the occupation numbers of $k\neq 0$-modes take only two possible values, $\langle \hat{a}_k^{\dagger}\hat{a}_k \rangle = 0$ or $1$, is 
   ${\mathcal N}_{states} = 2^{N_{modes}}$. The entropy resulting from such  micro-states scales as, 
    \begin{equation} 
      {\rm Entropy}  \, = \, \left ( {R \over L_{*}} \right )^{d-1} \, . 
   \label{ENTR}
   \end{equation}  
  By allowing the higher occupation numbers of the gapless modes,
  e.g., up to $\langle \hat{a}_k^{\dagger}\hat{a}_k \rangle <  n$, 
   we  can increase the entropy by a  log$(n)$ factor, but the area scaling 
   remains intact.  \\

Before concluding this section, we wish to make couple of comments. \\ 

First, when taking the limit in which some of the modes become gapless, 
we have to make sure that the limit is smooth. 
That is, the interaction terms must remain suppressed and be subdominant under the small excitations of the gapless modes.  
In other words, we must avoid a strong coupling problem. \\

The fact that in our case the limit is smooth,  should have 
already been clear from the fact that the double-scaling limit (\ref{doublescale}) allows to eliminate  interaction terms, prior 
 to taking the limit  (\ref{KMAXlimit}), which diminishes 
the gap $\delta_*$  of the $k_*$-modes, without restoring the interaction terms in  (\ref{terms}). 

In other words, since the gap $\delta_*$  and the interaction terms  are controlled by different parameters, for an {\it arbitrarily small} 
value of $\delta_*$, we can  make the interaction terms
{\it arbitrarily weak} by taking $N_0$ large enough (equivalently, 
taking $g$ small enough).  \\     
   
   Secondly, let us remark, that although according to the Hamiltonian 
 (\ref{HALSTAR}) all the degenerate micro-states have zero energies, this should not create a false impression that we manage to store 
 an unlimited information in the system, without paying an over-all 
 energy price.   We must remember that the macro-state 
 described by the large occupation number $N_0$ of the zero angular-momentum quanta, 
 costs  an ``inert" rest mass energy $E_0 = N_0mc^2$, where $c$ is the
 speed of light.  In non-relativistic limit this gives an overall infinite constant. However, this constant is invariant under a redistribution of  the total particle number $N_0$ among the different  angular-momentum-modes, and therefore, plays no role in
 the micro-state count.  So, the criticality
of the level $\epsilon_*$ is not eliminating this universal energy price that {\it any} system (including a black hole) must pay for achieving a large micro-state entropy. 

Nevertheless, the criticality phenomenon allows us to gain enormously in the 
density of states, i.e., fitting an exponentially large number of states within  an infinitesimal energy gap, thanks to a large number of the emergent gapless modes.  
 
 This situation is analogous  to what happens in black holes. 
 There too, 
 a large micro-state entropy of a black hole demands a huge  over-all 
 energy price in form of the rest mass energy of the black hole.
 In other words, having many degenerate black hole states
 is only possible if each of them is very heavy.

   \section{$S_3$ case} 
   
  Due to an obvious phenomenological importance of a three-dimensional 
  space, we shall present an explicit counting for $d=3$.  The  angular 
  coordinates $\theta^a$ 
  on a $3$-sphere can be taken as  $(\chi, \theta,\phi)$, where 
  $0 \leqslant  \chi \leqslant  \pi,  ~ 0  \leqslant  \theta  \leqslant \pi, 
   ~0\leqslant  \phi  \leqslant 2\pi$. The  interval and the volume elements  
   are given by  
   $ds^2 = d\chi^2 + {\rm sin}^2\chi (d\theta^2 + {\rm sin}^2\theta d\phi^2)$ and 
   $d^3\Omega = {\rm sin}^2\chi d\chi {\rm sin} \theta d\theta d\phi$, respectively. The spherical harmonics $Y_{klm} (\chi ,\theta, \phi)$ are  labeled by the three integers, satisfying the condition: $|m| \leqslant l \leqslant k =0,1,2,...\infty$.   
   These functions are the eigenmodes of the Laplace operator on an unit $S_3$, with the eigenstate equation given by, 
   $\Delta_{3} Y_{klm} = - k(k+2) Y_{klm} $. As usual, they form a complete orthonormal set. Now, using the properties of $Y_{klm}$-functions and 
 repeating all the above computations for $d=3$, we arrive  to the following 
 form of the Hamiltonian  in the double-scaling limit:
     \begin{eqnarray}
 &&\hat{H}  =  \\
=  &&  \sum_{|m| \leqslant l \leqslant k =1}^{k_{*}} \epsilon_k\Big\{
 1 - g N_0 \epsilon_k \Big\} \hat{a}_{klm}^{\dagger}\hat{a}_{klm} \nonumber \, .
 \label{HAL3} 
 \end{eqnarray}
 where $\epsilon_k  \equiv {\hbar^2  \over 2mR^2} k(k + 2)$. \\

   Again, taking the limit 
   \begin{equation} 
 \delta_* \equiv \epsilon_* \Big\{1 - g N_0 \epsilon_* \Big\} \rightarrow 0^+\,,
 \label{THElimit}
 \end{equation} 
 we get the 
 spectrum of the gapless modes $\hat{a}_{klm}^{\dagger}, \hat{a}_{klm}$ 
  with $k = k_{*}$. Their number is
 given by the corresponding degeneracy of the eigenvalue $k_*$ and is equal to,   
   \begin{equation} 
      N_{modes} \, = \, k_{*} (k_{*} + 2) \, = \,   \left ( {R \over L_{*}} \right )^{2} \, . 
   \label{Nmodes2}
   \end{equation}  
Thus,  the micro-state entropy scales as an area of a two-sphere: 
      \begin{equation} 
      {\rm Entropy}  \, = \, \left ( {R \over L_{*}} \right )^{2} \, . 
   \label{EN9}
   \end{equation} \\
   
  Notice, the gapless modes 
  $\hat{a}_{k_{*} lm}^{\dagger}, \hat{a}_{k_{*} lm}$ satisfy all the requirements for being called the {\it holographic}  degrees of freedom.
  In particular, they saturate almost  the entire information storage capacity 
  of the system. The quantum messages can be encoded into the basic states 
 with different  occupation numbers of these modes, 
 $\ket{0,0,...0}, \, \ket{1,0,...0}, \, \ket{0,1,...0}, \, ...., \ket{1,1,...1},\,...$, as well as in their superpositions.  The superpositions formed by these states can be highly 
 entangled and have a very high complexity.   
 
 \section{Bounded Hamiltonian} 
 
 In order to avoid creating a wrong impression as if the emergence of the area law relies on a truncation of the angular-momentum tower, we shall consider an example of the bounded-from-below Hamiltonian, for which no angular-momentum cutoff is required 
 for the stability. \\
  
 We choose the Hamiltonian in the following form: 
  \begin{eqnarray} 
  &&  \hat{H}  =  \int d^d\Omega \Big\{ - 
   \hat{\psi}^{\dagger} \Delta  \hat{\psi}\,  -  \\ \nonumber
   && - \, g \Omega \,  (\hat{\psi}^{\dagger}  \Delta  \hat{\psi}^{\dagger}) (\hat{\psi} \Delta  \hat{\psi}) \, +
 \tilde{g} \Omega (\hat{\psi}^{\dagger}  \Delta^2  \hat{\psi}^{\dagger}) (\hat{\psi} \Delta^2  \hat{\psi}) 
 \Big\} \,    
   \label{BH} 
\end{eqnarray}
where $ \tilde{g}$ is a new positive constant, of dimensionality of inverse energy-cubed.  
  Again, this Hamiltonian possesses a quantum critical point at which a large number of gapless 
modes emerge, resulting in the area-law micro-state entropy. \\ 

 In order to see this,  we again integrate over the space coordinates and take 
the double-scaling limit 
  \begin{eqnarray} 
 &&N_0 \rightarrow \infty\,, ~ ~  g  \rightarrow 0 \,, 
~~  \tilde{g}  \rightarrow 0 \\ \nonumber
 && gN_0 = {\rm finite}, ~~\tilde{g} N_0 = {\rm finite} \,.
 \label{doublescale2}
 \end{eqnarray}
  
The resulting 
 Hamiltonian has the following form: 
 \begin{eqnarray}  \label{Hstable} 
 &&\hat{H}  =  \\
=  &&  \sum_{k\neq 0}^{\infty}  \epsilon_k \Big\{
 1 - g N_0 \epsilon_k + \tilde{g}N_0 \epsilon_k^3\Big\} \hat{a}_k^{\dagger}\hat{a}_k \nonumber \, .
 \end{eqnarray}
 
 The difference from the previous case is that here we are left with 
 two control parameters: $g N_0$ and $\tilde{g}N_0$.
 From the above expression it is clear that the boundedness of the hamiltonian no longer 
 requires a cutoff, since 
 the modes with $\epsilon_k \rightarrow \infty$ are stable.\\

We shall now impose the constraint on the Hamiltonian (\ref{Hstable}) to be 
semi-positive definite.  Then, the  quantum criticality is reached
whenever the Hamiltonian touches zero for some $\epsilon_k = \epsilon_*$.   
Such a situation is achieved when the two control parameters, $(gN_0)$ and 
$(\tilde{g}N_0)$, take the following values: 
 \begin{equation}
 (gN_0) = {3\over 2} {1\over \epsilon_*}\,, ~~
 (\tilde{g}N_0) = {1\over 2} {1\over \epsilon_*^3}\,.
 \label{LLL}
 \end{equation} 
 With the above choice, the eigenvalues of the Hamiltonian (\ref{Hstable})
 are semi-positive definite and touch zero for $\epsilon_k = \epsilon_*$.
  This can be easily visualized by using (\ref{LLL}) and rewriting the Hamiltonian (\ref{Hstable}) in the form
  \begin{eqnarray}  \label{HAMILTON} 
 &&\hat{H}  =  \\
=  &&  \sum_{k\neq 0}^{\infty}  {\epsilon_k \over 2} 
\left({\epsilon_k\over \epsilon_*} -1\right )^2 \left({\epsilon_k\over \epsilon_*} +2\right )  \hat{a}_k^{\dagger}\hat{a}_k \nonumber \, .
 \end{eqnarray}
 In this case, only the modes with $\epsilon_k = \epsilon_*$, become gapless, while all the other modes 
 $\epsilon_k \neq \epsilon_*$ have the positive energy gaps and are stable. \\
 
 Thus, once again, the choice of the control parameters given by 
(\ref{LLL}), defines a one parameter family of theories
parameterized by the  scale of criticality $\epsilon_*$.  
Notice, the critical level $\epsilon_*$ is no longer a cutoff of the tower, but an arbitrary marker that separates the modes into the high-level  
({\it microscopic}) and the low-level ({\it macroscopic}) ones. 
Its value  can be chosen arbitrarily.  
 In this sense, the corresponding length-scale 
$L_* \equiv {\hbar \over \sqrt{2m\epsilon_*}}$, defined earlier, 
 plays the role of a non-gravitational ``Planck length" for this toy non-relativistic model. \footnote{Of course, the true gravitational 
 Planck length vanishes in non-relativistic limit.}   
\\

 Of course, it is natural to assume that $\epsilon_*$  is much higher  than the unit gap 
 ${\hbar^2 \over 2mR^2}$.    
 Then, repeating the previous counting, we immediately obtain that 
 the number of the gapless modes and  the entropy of the resulting micro-states are given by the equations 
 (\ref{Nmodes}) and  (\ref{ENTR}) respectively. Both of these quantities scale as the area of a $d-1$-dimensional sphere of radius $R$. \\

\section{Summary of the mechanism} 

 We  are now in the position to summarize the essence of the considered  mechanism. 
It consist of two crucial ingredients. These are:   {\it 1) The large-$N_0$ Criticality};  and {\it 2) the level-degeneracy due to the spherical symmetry}. \\

 Let us elaborate on the above points.  \\

  First of all, the high occupation numbers of some of the low-level modes - due to their attractive couplings to the 
 high-level ones - diminish the gaps of the latter modes.  Indeed, as we saw, for a high occupation number
 $N_0$ of the $0$-level mode and for a certain value of the criticality parameter,  the effective energy gaps of $k_*$-level 
  modes
 were lowered to zero. Notice,  in the state $N_0=0$, the  energy gaps for the same $k_*$-modes would be enormous, $=\epsilon_*$. 
 However, in the state in which $(k=0)$-level mode is critically occupied, the gaps of $k_*$-modes collapse to zero. \\
 
  {\it In a nutshell, a critical occupation of the low-level modes 
 collapses the gaps of the high-level ones. }  \\ 
 
  Once this is appreciated, the entropy counting follows solely from the 
  degeneracy of the level $\epsilon_*$, which is fully controlled by the 
  angular momentum conservation and the spherical symmetry of the problem. \\

   In is natural to hypothesize that the above phenomenon is indeed taking place in black holes, provided we assume that the black holes are the states in which some of the 
  soft gravitons are macroscopically occupied to a critical level \cite{DGN,DG}. 
  Then, the high occupation number of the soft gravitons, 
   suppresses the energy gaps of the Planck wavelength ones, resulting into the area law entropy. \\    
  
{\it Notice: This is a trade-off. By putting a lot of energy in high occupation number of soft gravitons, the black hole makes the excitations of Planck-wavelength  gravitons very cheap. }

\section{Discussion} 

 We have shown that a simple non-relativistic quantum system
 living on a $d$-dimensional sphere and describing 
 a bosonic field with an attractive self-coupling, possesses a quantum critical state with an exponentially enhanced 
 degeneracy of micro-states.  Remarkably, the entropy of these 
 micro-states is  given by an area of a $d-1$-dimensional sphere. \\
 
  {\it We must note, that thanks to double-scaling limit, the entropy-counting is exact and is not corrected by quantum fluctuations. 
 Of course, once we depart from this limit, we expect the finite gaps 
   to develop. However, they will be suppressed by ${1\over N_0}$ and thus will be under control in large-$N_0$ regime. } \\

   This model also gives a precise meaning to the concept of 
   holography. Indeed, the gapless modes that emerge at the critical point,  
  by all measures, represent the {\it  holographic}  degrees of freedom, with their number scaling as area of a sphere, of one dimension lower 
  than the bulk space.  \\

  How can we interpret the above result? 
 There are at least three different points of view that one could take. \\
   
 The first view is that the above quantum system does capture some 
 fundamental properties of quantum gravity, and therefore, the 
 area law of entropy is of {\it gravitational type}.  In particular, this point of view is supported by the fact that  the self-interaction of the bosonic field is both attractive and momentum-dependent, something that is highly reminiscent of gravity.  \\
 
  The second view is that the phenomenon of the enhancement of entropy and its  scaling according to the area law, goes well beyond gravity and we are observing one such example.  \\
 
  These two viewpoints are not inter-exclusive and they are the two components of the hypothesis of \cite{DG}. This is also our point of view. 
   That is,  we think that the area law scaling of 
 Bekenstein entropy originates  from quantum critical state reached due to the high occupation number of attractive gravitons.  However, in the same time, there can exist other systems of nature that can exhibit a similar behaviour.  \\
  
  Finally, there is a third view stating that the above-observed 
  area-law-type entropy  is just a remarkable  coincidence, unrelated to the black hole physics. This is a fully legitimate  point of view, although not our favourite one.  However, if it were true, it could teach us an equally important lesson. 
  This lesson would be that black holes may not possess a complete  monopoly on holographic scaling of entropy.  \\

\section*{Acknowledgements}

It is a pleasure to thank Lasha Berezhiani, Cesar Gomez and Sebastian Zell for discussions. 
This work was supported in part by the Humboldt Foundation under Humboldt Professorship Award, ERC Advanced Grant 339169 "Selfcompletion", by TR 33 "The Dark Universe", and by the DFG cluster of excellence "Origin and Structure of the Universe". 

\appendix

\end{document}